# Acoustic topological Anderson insulators

Hui Liu[1], Boyang Xie[1], Haonan Wang[1], Wenwei Liu[1], Zhancheng Li[1], Hua Cheng[1,*], Jianguo Tian[1], Zhengyou Liu[2,*], and Shuqi Chen[1,3,4,*]

[1]*The Key Laboratory of Weak Light Nonlinear Photonics, Ministry of Education, Renewable Energy Conversion and Storage Center, School of Physics and TEDA Institute of Applied Physics, Nankai University, Tianjin 300071, China*

[2]*Key Laboratory of Artificial Micro- and Nanostructures of Ministry of Education and School of Physics and Technology, Wuhan University, Wuhan 430072, China*

[3]*The Collaborative Innovation Center of Extreme Optics, Shanxi University, Taiyuan, Shanxi 030006, China*

[4]*The Collaborative Innovation Center of Light Manipulations and Applications, Shandong Normal University, Jinan 250358, China*

*Corresponding author, email: hcheng@nankai.edu.cn; zyliu@whu.edu.cn; schen@nankai.edu.cn

**Abstract:**

**Recent breakthrough on topological Anderson insulators revealed the breakdown of the traditional perception that sufficiently strong disorder may induce the appearance of topological protected transport states instead of destruction. Although topological Anderson insulators have been observed in various time-reversal symmetry breaking systems, the observation of topological Anderson insulators protected by time-reversal symmetry remains scarce, which are considered to be more promising in applications such as the integrated devices. Here, we report the experimental observation of topological Anderson insulator in a two-dimensional bilayer phononic crystal. The robust spin-dependent edge**

**states, as evidence of topological Anderson insulating phase, are observed by introducing on-site disorder. In addition, spin Bott index was computed to identify the topological invariants of the system with disorder, which confirmed the occurrence of disorder-induced topological state. Our results reveal that the impurities and defects introduced in the processing of integrated devices may induce the formation of topological transport states, which are promising for the exploration of new routes for the integration devices design.**

Topological insulators (TIs) have been an ongoing fascination in recent years as their topological protected edge states are ideal propagation states that are not found in the traditional systems, which may bring abundant changes in future devices. TIs, featuring topological gapless surface states arising from nontrivial topological origin of bulk states, have been predicted and realized in many periodic wave systems, including condensed matter[1,2], optics[3–5], acoustics[6,7], mechanics[8,9] and cold atoms[10,11]. By now, almost all the previous researches of the traditional TIs are based on periodic systems. Disorder is well known to be the obstacle in the traditional transmission, which can lead to the backscattering or localization of the transmission, known as Anderson localization[12–14]. In general, for TIs, whose robustness against weak perturbations is at heart of topological phenomenon, their topological protected transmission still cannot survive from the destruction of sufficiently strong disorder. However, a contrary proposal, known as topological Anderson insulator (TAI), indicates that the sufficiently strong disorder may push the realization of topological phase rather than inhibit, which stirs the interest of the researchers[15–18].

TAI was first predicted to occur in two-dimensional (2D) HgTe/CdTe quantum well in condensed matter[19–22]. Study found a disorder-driven topological phase show the transition from a trivial phase to a topological phase marked by the appearance of quantized conductance. Over the past decade, TAIs have drawn enormous interest and many theoretical works found the fact that they generally exist in various topological systems, including 2D time-reversal symmetry (TRS) breaking system[23,24], 2D quantum spin Hall effect (QSHE)[17,25–27] and three-dimensional (3D) QSHE[28,29]. However, two great difficulties were encountered during the research leading to few TAI experimental observations. One is the lacking of real materials in various systems where precise controlling of on-site potential energy and coupling strength face challenges. The other is the difficulty of quantifying topological invariants in aperiodic systems where the energy band theory and the topological invariants defined in the momentum space will become invalid, such as Z and $Z_2$ invariants. The breakthrough of TAI observations first occurred in Z insulators with TRS breaking, including one-dimensional (1D) cold atom wires[30], 2D photonic Floquet TIs[31] and 2D photonic Chern insulator[32]. Very recently, other disorder-driven phenomena, such as the 1D topological Anderson phase[33] and the higher-order TAI[34], are observed in 1D acoustic signal filtering and electric circuit respectively. Remarkably, most of these 2D topological Anderson transmission observed previously are based on the breaking of TRS, whose realization relies on external magnetic field or extra time control. TAIs protected by TRS have been long-predicted in $Z_2$ insulators, which have greater development potential in applications and integrated devices, and have yet to be observed due to the lack of ideal material.

In this work, we report an experimental realizations of TAI in a 2D TRS-preserving bilayer acoustic phononic crystal, which was previously found to give rise to the spin Chern insulator[7]. The disorder is introduced by randomly rotating the scatterers in each unit of bilayer phononic crystal. Starting with a trivial insulating phase of the phononic crystal with zero $Z_2$ invariant, we observed the occurrence of a pair of helical edge states when the disorder is sufficiently large, which is the hallmark of the topological Anderson phase transition. Such a pair of helical edge states are robust to obstacle, similar to the topological edge modes in spin Chern insulator, and are validated numerically and experimentally. In this TAI, the topological invariant in the system with spin nonconservation is described by spin Bott index instead of $Z_2$ invariant and the associated phase diagram further verifies the occurrence of disorder-driven topological phase transition in our system. Our work may open a door for applications on the disorder-induced transmission in various systems.

**Tight-binding model for TAI**

To illustrate the design philosophy of TAI in the acoustic system protected by TRS, we start with the tight-binding model in bilayer Lieb lattice where three type of atoms, denoted A (red sphere), B (blue sphere) and C (green sphere), are located at sublattice sites in each layer, as shown in Fig. 1a. We consider the system with pseudospin of the upper and lower layers denoted by subscript $\mu \in \{-1,1\}$ and the corresponding Hamiltonian is given by:

$$H = \sum_{i,\mu} U_{i,\mu} c^\dagger_{i,\mu} c_{i,\mu} + t_0 \sum_{\langle i,j \rangle,\mu} c^\dagger_{i,\mu} c_{j,\mu} + t_c \sum_{\langle\langle i,j \rangle\rangle,\mu} \{[\mu(\mathbf{n}_{ij} \cdot \mathbf{e}_z)+1]/2\} c^\dagger_{i,\mu} c_{j,-\mu}, \quad (1)$$

where $c^\dagger_{i,\mu} = (c^\dagger_{i,\mu}, c^\dagger_{i,-\mu})$ are creation operators on the up or down layer at the $i$ th site. The first term is the on-site energy $U_{i,\mu}$ for A, B and C atoms and the magnitude of which are expressed as the size of the red, green and blue spheres, as shown in Fig. 1a. The second term is the intralayer nearest-neighbor coupling with strength $t_0$. The last term describes the chiral interlayer coupling connecting two next-nearest-neighbor sites ($i$ and $j$) with strength $t_c$, where $\mathbf{n}_{ij} = \mathbf{e}_{ik} \times \mathbf{e}_{kj}$ and $k$ is the unique intermediate site of $i$ and $j$. Here, the double layers contribute to a layer degree of freedom acting as pseudospin and the chiral interlayer coupling introduces a synthetic spin-orbit coupling in the system resulting in the appearance of two topologically nontrivial bandgaps. The spin operator of $H$ is $\tau_y = \sigma_y \otimes I_N$, where $\sigma_y$ act as the layer degree of freedom and $N$ is the number of intralayer atoms. In this system, $\tau_y$ and $H$ are not commutative, which means the pseudospin is not conserved. However, a pair of gapless helical edge states with pseudospin polarization rely only on the protected TRS. Spin conservation of the system can be relaxed, and spin Chern number is well defined in this model with spin nonconservation.

A schematic depiction of the mechanism for TAI is shown in Fig. 1a. The system is driven from a trivial phase to topological phase by adding on-site disorder in a supercell. By adding strong enough on-site disorder to three kinds of atoms denoted by random size of red, blue and green spheres, this model exhibits topological Anderson phase transition characterized by the emergence of edge modes with pseudospin up and down polarizations, which are indicated by red and blue arrows on the hard boundaries. Here, for further investigating the mechanism of topological Anderson phase and the disorder types that may induce its occurrence, we first consider the topological phase transition in the unit cell of bilayer Lieb lattice. We focus on the lower bandgap by setting $U_{i \in A} = -m$, $U_{i \in B} = U_{i \in C} = m$, whose topological properties originate from lower two bulk bands. The associated topological phase transitions as a function of $m$ and $t_c$ are shown in Fig. 1b. Three topologically distinct phases for lower two bands noted with three colored areas are described by spin Chern number $C_s = (C_+ - C_-)/2 = 1, 0, -1$. At the boundary of two topological phase, the lower bulk bandgap closes implying the intermediate states of the phase transition. The bulk structure at high-symmetric point M as the function of $m$, $t_0$ and $t_c$, and the associated complete topological phases are analyzed in Supplementary Note 1. It is notable that different from the parameters $m$ and $t_c$, the intralayer coupling $t_0$ is independent of topological phase transition, which means the failure of the realization of topological Anderson phase transition by applying random disorder on $t_0$. Considering a ribbon with $1 \times 21$ lattices, we compared two typical projected band structures of the model ($t_0 = -2$ and $t_c = -0.25$) by setting $m = 0.5$ for trivial phase ($C_s^I = 0$) and $m = 0$ for topological phase ($C_s^I = 1$) respectively, as shown in Fig. 1c,d. For the trivial case, the lower bandgap is trivial and does not support any edge modes. While for the topological case, a pair of edge states with pseudospin up and down polarizations locking with +k and –k in the momentum exist in the lower topological bandgap denoted by red and blue lines, respectively. This pair of edge states with opposite pseudospin polarizations has opposite group velocities distributing on one hard boundary of the ribbon while the other pair with localize at the other hard boundary. As the phase transition diagram shown, the topological properties of the system rely on the on-site energy of the A, B, C atoms and the chiral interlayer coupling $t_c$. Here, we consider the influence of on-site energy types on topological Anderson phase transition. Starting in a $21\times 21$ supercell with the trivial phase by setting $t_0=-2$, $t_c=-0.25$, $m=0.35$ denoted by green star in Fig. 1b, random on-site disorder is introduced into each atoms of the supercell by $U_{i\in A;\mu}=-m+\delta U$, $U_{i\in B,C;\mu}=m+\delta U$, where $\delta U=[-w/2,w/2]$ and $w$ is the disorder strength. As the disorder strength $w$ increases, a trivial gap may close and then reopen and this trivial system can transit into a non-trivial phase. As the band structures are not well-defined in a disordered system, we calculated the energy eigenvalues instead of band structure for TAI in the supercells with periodic boundary condition (PBC) and hard boundary condition (HBC) in Fig. 1e. The energy eigenvalues in the supercell with PBC present a gap denoted by cyan area. The eigenvalues with HBC further confirm the disorder-induced phase transition, which features the occurrence of topological edge states within the gap region from a trivial gap. The wave function distribution of a typical mid-gap state shows the character of boundary state, where wave function distribution is located on the hard boundary of the supercell (Supplementary Note 3 provides details). Theoretically, when the supercell is infinitely large, the eigenvalues with PBC can fill all the range of bulk states, meanwhile the eigenvalues with HBC can completely fill the topological band gap with edge states. The details of band structure for richer phase diagram and for distinct types of disorder adding to $m$ and $t_c$ respectively in the ribbon are presented in Supplementary Note 2. As the claim discussing in Supplementary Note 1,2, we propose that the disorder of on-site energy and chiral interlayer coupling can induce the topological Anderson phase transitions, while the intralayer coupling independent of the system phase transition cannot. The further discussions for three distinct types of $m$, $t_0$ and $t_c$ disorder in the supercells, including disorder-driven phase transitions, mid-gap eigenvalue and associated edge eigenstate, are shown in Supplementary Note 3.

In order to identify the process of topological Anderson phase transition, we need to verify the topological properties in a disordered system[35–38]. In the following, we calculate the topological invariant in terms of spin Bott index[39,40] instead of $Z_2$, which is invalid in aperiodic system. We analyze a 2D supercell of the size $L_x \times L_y = N_x \times N_y$ with PBC imposed in two spatial directions, where $N$ is a large enough unit cells number. We construct the projector operator of the eigenstates below the lowest gap:

$$P = \sum_{i}^{N_{occ}} |\psi_i\rangle\langle\psi_i|, \qquad (2)$$

where $N_{occ} = 2 \times N_x \times N_y$ is the occupied states below the lower gap. The next key step is to separate $N_x \times N_y$ pairs of eigenstates into the pseudospin up and down states. We project the projector operator ($P$) into the pseudospin space:

$$P_y = P\tau_y P, \qquad (3)$$

where $\tau_y = \sigma_y \otimes I_N$, $N = 3 \times N_x \times N_y$ is the number of atoms in a layer, $\sigma_y$ is the Pauli matrix. For our systems without spin conservation, $H$ is not commute with $\tau_y$ and $P_y$. The spin-mixing term, which breaks the spin conservation in $H$, depending on the chiral interlayer coupling $t_c$ is not strong, the eigenvalues of $P_y$ can be successfully split into two opposite groups ($\pm\hbar/2$) according to pseudospin up and down. Associated eigenvalue problem can be expressed as:

$$P_y|\pm\varphi_i\rangle = S_\pm|\pm\varphi_i\rangle, \tag{4}$$

where $S_\pm = \pm\hbar/2$. Here, we reconsider the projector operator of the eigenstates with pseudospin up and down, respectively:

$$P_\pm = \sum_i^{N_{occ}/2} |\pm\varphi_i\rangle\langle\pm\varphi_i|. \tag{5}$$

The corresponding projected position operators in $x$ and $y$ spatial directions for pseudospin up and down represent as:

$$U_\pm = P_\pm e^{i2\pi X} P_\pm, \tag{6}$$

$$V_\pm = P_\pm e^{i2\pi Y} P_\pm, \tag{7}$$

where $X$ and $Y$ are diagonal matrices with atomic coordinates in $x$ and $y$ spatial directions as diagonal elements. The spin Bott indices for pseudospin up and down are now given by:

$$B_\pm = \frac{1}{2\pi}\text{Im}\left\{tr\left[\log(\widetilde{V}_\pm \widetilde{U}_\pm \widetilde{V}_\pm^\dagger \widetilde{U}_\pm^\dagger)\right]\right\}. \tag{8}$$

Finally, the spin Bott index can be written $B_s = \frac{1}{2}(B_+ - B_-)$. It is noteworthy that this method relies on the existence of an absolute gap, without which the calculation of topological invariants may be inaccurate.

**2D bilayer acoustic TAI metamaterials**

We now consider the structural implementation of TAI with on-site energy disorder in acoustic spin Chern insulator. The air structure in the unit cell of phononic crystal sample is shown in Fig. 2a, consisting of two layers of air with height $h_1 = 8.4$ mm modulated by two square scatterers with width $l = 14.4$ mm and connected with four chiral tubes with diameter $d = 3.3$ mm and height $h_2 = 9.6$ mm. The square unit cell hosts the lattice constant $a = 24$ mm and the total height $h = 26.4$ mm. The pink and green square scatterers rotate in two air layers respectively, so as to the control of the bilayer on-site energy of ABC atoms and the intralayer coupling $t_0$. Top view of two scatterers is shown in Fig. 2b, represented with the pink square for the upper layer and the green square for the lower layer. The Angles between the diagonal of the square scatterers and the $y$ direction in the upper and lower layers are defined as $\theta_1=\theta_0+\theta_d R_i$ and $\theta_2=\theta_0+\theta_d R_j$, respectively, where $\theta_0$ is the initial scatterers rotation angle, $\theta_d$ is the modulation parameter characterizing the disorder strength of each layer, and $R_i$ ($R_j$) is a random number uniformly distributed between −0.5 and 0.5 for each unit cell in upper (lower) layer. Considering the disorder-free system by setting the disorder strength $\theta_d=0$ and the scatterers rotation angles $\theta_1=\theta_2=\theta_0$, we explore the corresponding bulk band structure in the unit cell for the periodic system. Since the unit cell is composed of two-layer ABC atoms, the bulk band dispersion of the model, which is performed by COMSOL Multiphysics, is composed of six band structures. The band structures at high symmetry points M in Brillouin zone degenerate in pairs and form three pairs of degenerate bands as a function of the angle of the scatterers $\theta_1=\theta_2=\theta_0$, shown in Fig. 2c. During the rotation of the scatterers, the upper bandgap closes at $\theta_0=\pm29°$ and the lower bandgap closes at $\theta_0=\pm10.5°$ where the phase transition and bands inversion occur. To obtain a clean bandgap, we focus on the phase transition in the lower bandgap among the lower and middle two bands, which closes at $\theta_0=\pm10.5°$. Two topological distinct phases, defined by spin Chern number of two lower bands, are shown by the pink area for trivial phase ($C_s^I=0$) and the blue area for topological phase ($C_s^I=1$) in Fig. 2c. To further investigate these distinct topologically phase, we explore the corresponding bulk band structures along high symmetry lines and the associated projected band structure for given scatterers rotations in Fig. 2d,f and Fig. 2e,g. A topologically trivial simulated bulk dispersion with $\theta_0 = 0°$ is plotted in Fig. 2d, where the topological trivial completely bandgap is indicated by pink area. As the scatterers rotation increases to $\theta_0 = 10.5°$, the topological trivial bandgap will close and a double Dirac cone occur at M point. For a further increase of $\theta_0 = 45°$, a completely bandgap reopen in a topological phase and the corresponding frequencies are denoted by blue area, shown in Fig. 2e. The appearance of the helical edge states in topological bandgap relies on the close and reopen of the double Dirac cone at M point. To demonstrate the topological properties of two distinct bandgap, we plot the projected band dispersions along $k_x$ for trivial and topological samples in Fig. 2f and Fig. 2g respectively, where the color maps are experimental dispersion obtained by Fourier transforming of the hard boundary excitation field and the overlaid lines are results from full-wave finite-element simulations. The excitation source is placed at one side of a hard boundary in the samples and this means only one edge state is design to excite. The experimental and simulated projection band dispersions for topologically trivial phase, acting as the initial phase of TAI, show a complete trivial bandgap in Fig. 2f. While the experimental and simulated projection band dispersions of TI show associated one-way edge modes in a topological bandgap in Fig. 2g. One edge state carrying pseudospin-up polarization is excited in the experiment shown in color map, which is consistent with the simulation result denoted by the red line. The other edge state with pseudospin-down polarization is not excited represented by simulation result with green dash line. In our experimental system, the topological phase transition caused by the rotation of scatterers actually results from the variation of atomic on-site energy. Although the variation of the intralayer coupling $t_0$ with the rotations of scatterers does not contribute to topological phase transition, it ensures a clean lower gap, which is beneficial to obtain a completely gap in a topological Anderson phase transition.

**Helical boundary states for topological Anderson phase transition**

Our experimental TAI sample with $21 \times 21$ unit cells was fabricated by 3D-printed structures shown in Fig. 3a, which constructed by two layers of Lieb lattices with rotated scatterers and a layer of interlayer connections with chiral tubes. The small holes are reserved in the upper layer for measure the acoustic field with the excitation at the hard boundary. In the TAI sample, the on-site disorder is introduced by two square scatterers with random rotation angles located at the upper and lower layers, respectively. Studying topological Anderson phase diagram is crucial to get more insights into the formation of the acoustic TAI, particularly regarding the transition of topological invariant with increasing of on-site disorder from trivial to nontrivial topology. We consider a supercell with random disorder $R_i$ and $R_j$ in the upper and lower layers and simulate the spin Bott index in a $11 \times 11$ unit cells with increasing disorder strength $\theta_d$ from the initial scatterers rotation angle of $\theta_0$. A schematic depiction of topological Anderson phase transform described by spin Bott index averaged over 10 times in the $\theta_0$-$R_i$ plane is shown in Fig. 3b, which confirms the inevitability of the formation of acoustic TAI with increasing disorders in our experimental system. In the experiment, three samples with $21 \times 21$ unit cells for trivial insulator, TAI and TI were fabricated, respectively. Two samples demonstrate the process of topological Anderson phase transition pointed by the arrow in the phase diagram, where the initial one is topological trivial phase ($B_s^I=0$) with $\theta_1=\theta_2=\theta_0=0°$ (pink region) and the final one is topological Anderson phase ($B_s^I=1$) with increasing random rotation Angle $\theta_1=\theta_0+R_i\theta_d$ and $\theta_2=\theta_0+R_j\theta_d$ ($R_i \neq R_j$, $\theta_0=0°$ and $\theta_d=180°$) in the upper and lower layers (blue region). In addition, the third sample with $\theta_1=\theta_2=\theta_0=45°$ for topological phase ($B_s^I=1$) is also measured for comparison. For the TAI case, we set the boundary of the sample as PBC and HBC, respectively. The associated simulated eigen frequencies for different state number are plotted in Fig. 3c, where the system with PBC shows an approximate topological gap range (7162 Hz-7537 Hz) denoted by blue area and the system with HBC shows several pairs of boundary states in the gap. For a larger supercell with $31\times31$ unit cells, more eigenvalues with PBC will fill the bulk frequency area and show a more accurate gap. More eigenvalues with HBC condition will fill this topological gap, as shown in Supplementary Note 5. The details of the phase transition and the formation of TAI with $21\times 21$ unit cells (case for $R_i \neq R_j, R_i=R_j$) with the increasing disorder strength in the TAI sample are shown in Supplementary Note 6. In addition to the transition of spin Bott index, the emergence of associated helical edge states, which is also the character in the TAI, can be observed with the mid-gap excitation at the hard boundary of the TAI sample. Hence, the edge states with pseudospin-up polarization of three samples were designed to excite at the frequency of 7300 Hz on a hard boundary. The simulated and observed acoustic fields for TAI case at the frequencies of 6700 Hz, 7300 Hz and 7900

Hz are shown in Fig. 3d-f and Fig. 3g-i, respectively. The acoustic fields at around the gap frequency (7300 Hz), indicated by the yellow dash line in Fig. 3c, show transmission support along the hard boundary for TAI case, which means the system is induced from a trivial phase to the topological phase by disorder. In addition, the localization of edge modes and the associated decay lengths are discussed in Supplementary Note 4. Both transmission and localization at gap frequencies demonstrate the existence of topologically protected edge states in TI and TAI samples.

**The backscattering robustness of boundary waves in TAI**

To gain further insights into the topological transmission properties of the TAI, we investigate the robustness of the pseudospin-dependent boundary states against a rectangular defect at the frequencies of 6700 Hz, 7300 Hz and 7900 Hz for the trivial insulator, TI and TAI in Fig. 4. These three frequencies correspond to the bulk modes at lower band dispersions, the modes in band gap and the bulk modes at higher band dispersions, respectively. The acoustic field are measured through the holes reserved in the upper layer of the sample. The intensity of acoustic field will radiate from the excitation source to the interior of the three samples excited at frequencies of 6700 Hz (Fig. 4a,d,g) and 7900 Hz (Fig. 4c,f,i), which exhibit bulk mode distribution. There are apparently distinguished acoustic field distributions in the topological trivial and nontrivial systems excited at mid-gap frequency of 7300 Hz. As there is no eigenmode in the trivial gap, the intensity of acoustic field is confined to the vicinity of the excitation source, as shown in Fig. 4b. While in the topological gap, the intensity of acoustic field is distributed at the hard boundary and propagates smoothly around the defect in Fig. 4e, which prove the existence of the topological protected boundary mode in the topological gap. In the case of TAI shown in Fig. 4h, the acoustic field distribution excited at 7300 Hz presents topological protected boundary mode similar to the topological one in Fig. 4e, but distinctly different from trivial case in Fig. 4b, which is the evidence of TAI's emergence. Similarly, another TAI sample with the same disorder in the upper and lower layers ($R_i = R_j$) is measured and shown in Supplementary Note 6. Our experiments show that the transmission of TAI through the defect is nearly unaffected, which means the backscattering is largely suppressed. In addition, for the TAI case, the Anderson localization can simultaneously occur with the addition of disorder, which may result in some transport states being localized or being difficult to excite experimentally.

In summary, we have observed disorder-driven topological spin-dependent edge states associated with the TAI in a 2D TRS-preserving acoustic system. The disorder-driven trivial-to-topological transition in aperiodic system is confirmed by calculating the spin Bott index for topological invariant. We also investigate three distinct types of disorder, added to the on-site energy, intralayer coupling and chiral interlayer coupling respectively, which may further prompt research on the role of disorder types in transmission. Our work is distinct from the earlier works on TAI, mainly reflected in the following aspects. Our experiments were conducted in an acoustic system that has little response to magnetic fields, verifying that the implementation of TAI does not depend on breaking the TRS, which may encourage the exploitation of disorder in integrated device applications. Furthermore, our experiments further demonstrate that the implementation of TAI is not limited to the systems with strong spin-orbit coupling or spin conservation condition, and that TAI can also be observed in spinless acoustic system with synthetic weak spin-orbit coupling and spin nonconservation, which may stimulate the explorations and applications of TAI in various systems.

**Data availability.** The data that support the finding of this study are available from the corresponding author upon request.

## Acknowledgements

This work was supported by the National Key Research and Development Program of China (2017YFA0303800), the National Natural Science Fund for Distinguished Young Scholar (11925403), the National Natural Science Foundation of China (11974193 and 91856101), and Natural Science Foundation of Tianjin for Distinguished Young Scientists (18JCJQJC45700).

## Author contributions

H.C., Z.L., and S.C. conceived the idea. H.L., B.X., H.W., and H.C. performed the theoretical analysis and numerical simulations and experiments. H.L., W.L., Z.C.L., H.C., J.T., Z.L., and S.C. prepared the manuscript. All the authors contributed to the analyses and discussions of the manuscript.

## Additional information

**Supplementary Information** is available online at XXX or from the author.

**Competing financial interests:** The authors declare no competing financial interests.

# Figures

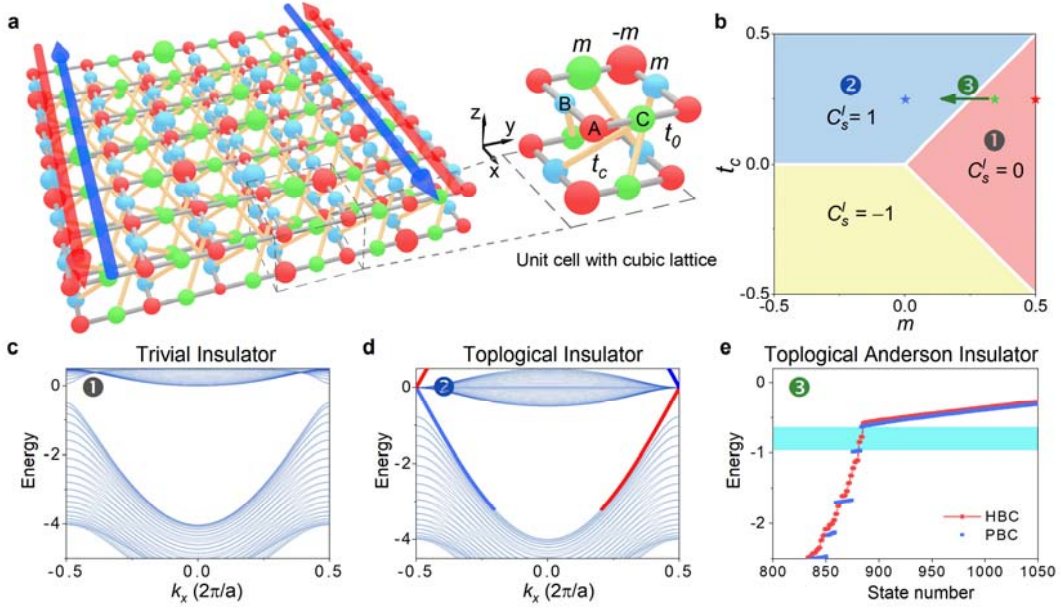

**Fig. 1 | Acoustic TAI in a bilayer Lieb lattice model**. **a**, Schematic illustration of Acoustic TAI in the presence of uniformly on-site energy disorder. The red, blue and green spheres of each layer indicate A, B, and C sublattices, and the variable radius represent the disorder on-site potential. Two pairs of topological edge modes with opposite pseudospins denoted by red and blue arrows localize at two hard boundaries respectively. **b**, A schematic phase diagram of unit cell distinguished by spin Chern number of the lower two bands in the $t_c - m$ plane. The white line represents the closure of the lower band gap, thus dividing the topological properties of the lower band gap. The green arrows indicate the process for on-site energy type of disorder-induced phase transition. **c**, Band structure for trivial insulator by taking $t_0 = -2$, $t_c = -0.25$, $m=0.5$ denoted by red star in **b**. **d**, Band structure for TI by taking $t_0 = -2$, $t_c = -0.25$, $m=0$ denoted by blue star in **b**. **e**, Calculated energy eigenvalues versus the state number for TAI in a $21 \times 21$ supercell with PBC (blue dots) and HBC (red dots). A sufficiently strong on-site energy disorder drives the system from the trivial phase to

topological phase and opens a topological bandgap showed by cyan area.

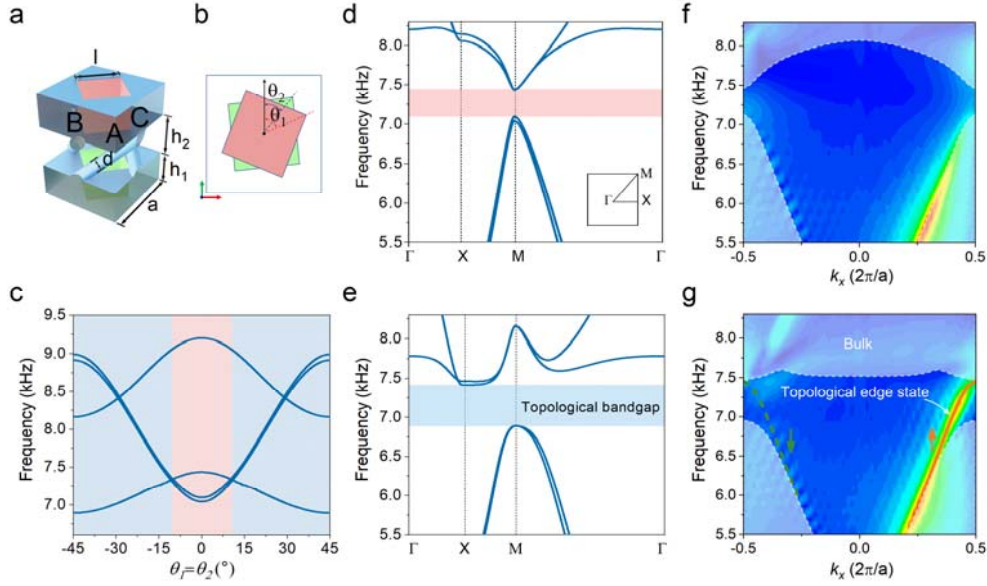

**Fig. 2 | The structural implementation of acoustic TAI and the band structure for two distinct disorder-free phases. a**, Unit cell of the disorder-free acoustic TAI. Pink and green cubes represent scatterers in the upper and lower layers, respectively. Three type of atoms are located at sublattice sites in each layers denoted by A, B and C. Air fills all the blue area. The gray boundaries along the *x* and *y* directions are PBC and the other boundaries are hard boundary covered with 3D printed material. **b**, Simplified top view of TAI cell structure. Two colored squares correspond to the scatterers in the upper and lower layers. Disorder is introduced by the random angles $\theta_1$ ($\theta_2$) of the upper (lower) scatterers uniformly distributed between $-\theta_d/2$ and $\theta_d/2$. **c**, The angular dependent frequencies for the unit cell with $\theta_1 = \theta_2 = \theta_0$ at high-symmetric point M. The trivial and the topological phase determined by the spin Chern number of the lower two bands are distinguished by pink ($C_s^I = 0$) and blue regions ($C_s^I = 1$), respectively. The simulated bulk band structure of **d** trivial insulator with rotational angles

$\theta_1 = \theta_2 = 0°$ and **e** TI with rotational angles $\theta_1 = \theta_2 = 45°$ along high symmetry lines. The pink and blue areas in **d** and **e** denote trivial and topological bandgaps, respectively. The projected band dispersions for **f** trivial insulator and **g** TI (only spin-up component is excited). The color maps represent the measured data. The shadow regions and colored line indicate the simulated data of projected bulk states and edge states.

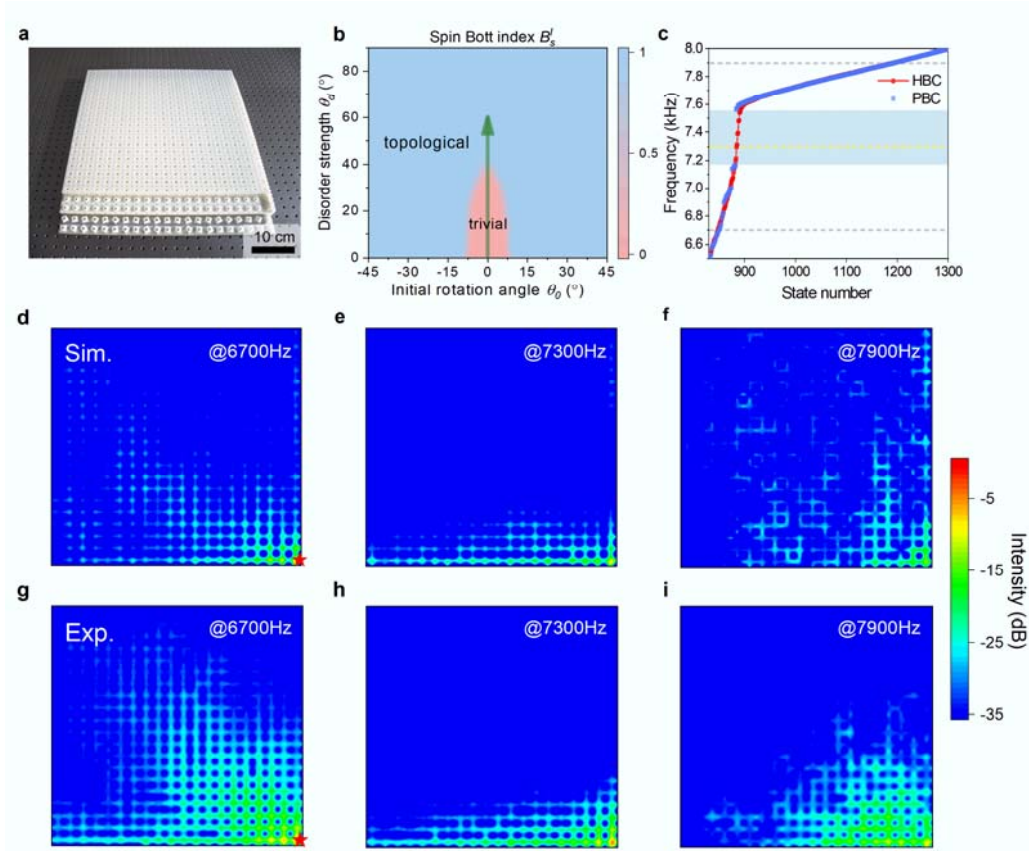

**Fig. 3 | Disorder-induced phase transition from trivial to topological phase. a**, A photo of TAI sample. **b**, Simulated TAI phase diagram showing the average spin Bott index as a function of the initial rotation angle and the disorder strength simulated in a 11×11 lattices containing 726 atoms. **c**, Simulated eigenfrequency for TAI in a 21×21 supercell with PBC (blue dots) and HBC (red dots). The gap frequencies (general range: 7162 Hz-7537 Hz) are shown in blue area and excitation frequencies are indicated by yellow dash line. **d-f** Simulated and **g-i** experimental boundary states excitation acoustic fields for TAI in a 21×21 lattices at the frequency of 7600 Hz, 7300 Hz and 7900 Hz corresponds to lower bulk mode, edge mode and upper bulk mode, respectively. At the frequency (7300 Hz) in the gap, TAI shows the exist of edge

transmission modes on the hard boundary and differs from the initial trivial case, whose transmission is suppressed.

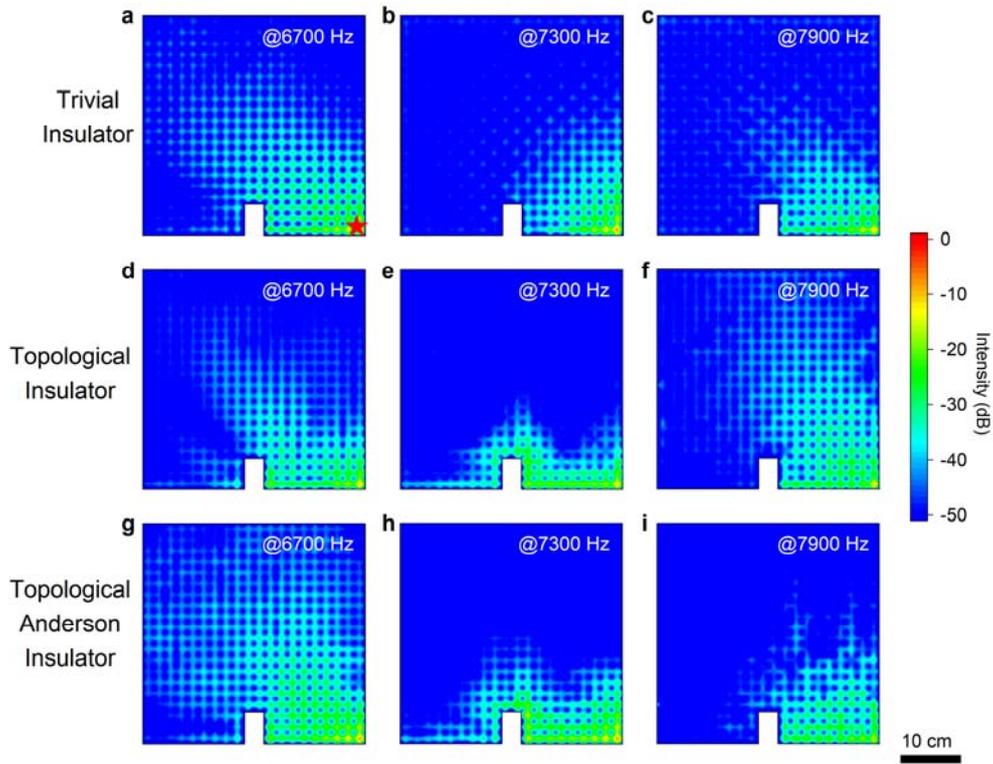

**Fig. 4 | Robustness of pseudospin-dependent edge states against defect in TAI compared with trivial insulator and TI.** The field distributions of **a-c** trivial insulator, **d-f** TI and **g-i** TAI excited at the frequency of 7300 Hz, 6700 Hz and 7900 Hz on the hard boundary, where red star represents excitation source. Among them, 7300 Hz corresponds to mid-gap frequency, and the other two frequencies correspond to higher frequency and lower frequency bulk states.